\documentclass[cits]{PoS}

\usepackage{verbatim}
\usepackage{amsthm}
\usepackage{amsmath}

\newcommand{\harxold}[1]{\href{http://arxiv.org/abs/hep-lat/#1}{\tt 
hep-lat/#1}}
\newcommand{\harxnew}[1]{\href{http://arxiv.org/abs/#1}{\tt #1}}
\newcommand{\hippy}{\texttt{HiPPy}}
\newcommand{\hpsrc}{\texttt{HPsrc}}
\newcommand{\hipsrc}{\texttt{HiPPy}/\texttt{HPsrc}}
\newcommand{\pastor}{\texttt{Pastor}}
\newcommand{\physycal}{\texttt{PhySyHCAl}}

\title{Automated Lattice Perturbation Theory}

\ShortTitle{Lattice Perturbation Theory}

\author{\speaker{Christopher Monahan}\\ College of William and Mary\\
E-mail: \email{cjmonahan@wm.edu}}

\abstract{I review recent developments in automated lattice perturbation
theory. Starting with an overview of lattice perturbation theory, I 
focus on the three automation packages currently ``on the market'': 
\hipsrc{}, \pastor{} and \physycal. I highlight some 
recent applications of these methods, particularly in $B$ physics. In the 
final section I briefly discuss the related, but distinct, approach of 
numerical stochastic perturbation theory.}

\FullConference{31st International Symposium on Lattice Field Theory -
LATTICE 2013\\ July 29 - August 3, 2013\\ Mainz, Germany}

\begin{document}

\section{Introduction}

The lattice is generally regarded as a nonperturbative tool. 
This does not mean, however,
that perturbative analyses of lattice field theories are redundant.
Many of the early calculations in lattice field theory incorporated analytic 
treatments of
weak coupling behaviour, because initial efforts were often constrained by 
insufficient computational
power. As 
lattice field theory matured, increasing computing power and improved 
algorithms facilitated fully nonperturbative treatments with greater 
precision. Parallel theoretical developments, such as the clarification 
of conceptual issues with the pole mass, often demonstrated the need for fully 
nonperturbative treatments, particularly when aiming for high precision

Nevertheless, lattice perturbation theory, which is perhaps better thought 
of as ``perturbation theory for lattice actions'', has retained an important 
role for a variety of lattice calculation. Many of these applications fall into 
one of three general categories (see, for example, the references collected in 
\cite{lptorg}): determining the renormalization parameters of bare lattice
actions; matching regularisation schemes and extracting continuum
results from lattice data; and improving lattice actions. These
categories are inter-related and can be broadly construed
as accounting for the effects of energy excluded by the lattice
cutoff. In other words, LPT calculations are generally 
concerned with renormalisation.

In the next section I discuss the motivation and justification for LPT and in 
Section \ref{sec:auto} I review automated lattice perturbation theory, 
focussing on the three automation packages currently available and highlighting 
some applications of these techniques. In Section \ref{sec:nspt} I discuss the 
related, but distinct, approach of numerical stochastic perturbation theory.

\section{Perturbative and nonperturbative renormalisation}
\label{sec:renorm}

The
lattice is a gauge-invariant ultraviolet regulator that, by discretising 
spacetime, excludes all momenta greater than $\pi/a$ (where $a$ is the lattice
spacing). To correctly account for energy scales eliminated by the cutoff, 
the regularised theory must be renormalised. In continuum gauge theories, one 
is generally restricted to perturbative renormalisation, but on the lattice 
both perturbative and nonperturbative techniques are available. 

There is a wide variety
of nonperturbative methods and a full discussion is beyond the scope of
these proceedings, but a partial list includes nonperturbative tuning via 
physical quantities or dispersion
relations (see, for example, \cite{aoki2012}); imposing Ward identities or 
chiral
symmetry relations \cite{luescher1996}; and step-scaling methods, which have 
been applied to the Schr\"odinger functional \cite{luescher1991}, off-shell
Green functions (the ``Rome-Southampton method'') \cite{martinelli1995},
physical quantities \cite{lin2007} and gradient flow smearing procedures 
\cite{orginos2012}. 

Although different in practice, nonperturbative methods share
two distinct advantages. The first is that nonperturbative techniques 
eliminate perturbative truncation errors, which can be hard to
quantify, in favour of statistical and systematic errors, which can generally 
be reliably
determined and, with intensified computational effort,
systematically reduced. The second is more aesthetic: only a fully
nonperturbative approach is truly \emph{ab initio}.

Unfortunately there are disadvantages:
step-scaling and iterative methods incur a greatly increased
computational expense, matching to physical quantities results in
a loss of predictive power and some methods explicitly 
break gauge invariance.

Lattice perturbation theory (LPT) is often an attractive alternative, 
computationally cheaper and without loss of predictive power. Naturally there is 
a cost. LPT introduces perturbative truncation errors and as lattice
calculations become more precise, the perturbative errors
often become the largest source of uncertainty in the final result. 
Furthermore, reliably estimating higher-order truncation errors can be 
difficult. In many cases, though, the use of LPT is justified by a 
consideration of the relevant energy scales in question. For current lattice 
calculations the scales excluded by the inverse lattice spacing are those above 
(approximately) 5 GeV. At such scales, the coupling constant is sufficiently 
small, $\alpha_s(\pi/a)\sim
0.2$, that perturbative approximations to the renormalisation parameters are likely to
be valid. 

LPT therefore connects the low and high energy regimes
of a lattice field theory. For lattice quantum chromodynamics (QCD) this 
connection has been tested and validated in a
wide range of processes by comparing perturbative
calculations with polynomial fits (in the coupling constant) of 
nonperturbative data in the weak coupling
regime
\cite{lepage2000,direnzo2001,horsley2002,trottier2002,wong2006,
allison2008}.

This isn't quite the whole story, however. Early LPT calculations were plagued 
by slow convergence
and inconsistent results \cite{lepage1993}. These issues were the consequence
of a poor choice of expansion parameter: the bare lattice 
coupling $\alpha_\mathrm{latt} = g_0^2/(4\pi)$. The convergence 
of the perturbative series can be improved by
introducing a different coupling constant \cite{lepage1993}. For example, one 
common choice of scheme for lattice QCD is the
``V-scheme'' \cite{brodsky1983}, expressed at an appropriately chosen
scale, usually the ``BLM scale'' \cite{brodsky1983,hornbostel2002}, that 
characterises the scale of the process concerned. The convergence of the 
perturbative expansion can be enhanced by tadpole improvement 
\cite{lepage1993}, which partially mitigates the effects of high 
momenta contributions from so-called tadpole diagrams. This improvement has now
been generalised to remove ``cactus'' diagrams \cite{cps2006}, although this 
technique is less widely used.

The main challenge for perturbative renormalisation is the complexity of 
loop calculations in LPT. Even for the simplest lattice actions, the use of 
gauge links rather than gauge fields increases the number of Feynman diagrams 
that must be evaluated at a given order by generating vertices with an 
arbitrary number of gauge fields (consistent with the order of the calculation, 
of course).

For many modern lattice actions, however, things are worse. 
Sophisticated lattice actions often include irrelevant operators, smeared gauge 
links to improve discretisation effects, or tadpole (or cactus) improvement, and 
these 
considerably complicate the Feynman rules.
Highly improved actions, such as the highly-improved staggered quark (HISQ) and 
nonrelativistic QCD (NRQCD) actions, generate Feynman rules that cannot
feasibly be manipulated by hand. Furthermore, the lack of Lorentz
symmetry ensures that Feynman integrands are no
longer amenable to the same tricks and transformations that are available 
for continuum theories. To cope with all of
these complications LPT has been automated.

\begin{figure}
\begin{center}
\hspace*{-30pt}\includegraphics[width=0.5\textwidth,keepaspectratio=true]
{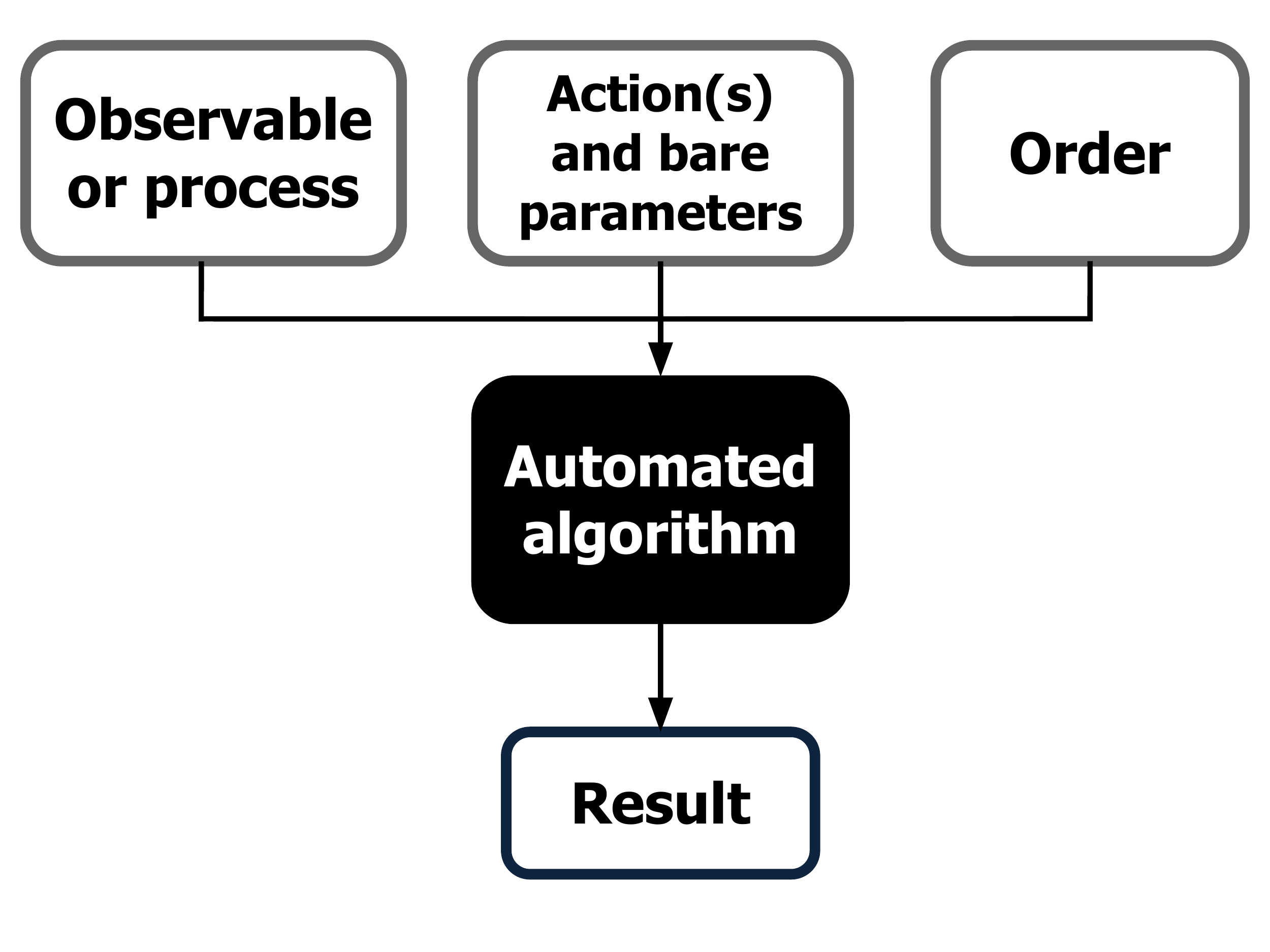}
\end{center}
\caption{Flowchart illustrating an idealised, ``black box'' automation 
process.\label{fig:auto}}
\end{figure}

Before proceeding, it is worth considering what ``automation'' 
involves in the context of LPT. A formal definition is not 
necessary, but to convey a general sense of what I mean by ``automation'' I 
represent a paradigm example of an ALPT 
algorithm in Figure \ref{fig:auto}. The extent to 
which the user must fill in the steps represented by the ``black box'' at the 
heart of the process provides a heuristic measure of the level of automation 
of 
a particular routine.

\section{Automated routines}
\label{sec:auto}

The dawn of ALPT dates
to the 1980s, when L\"uscher and Weisz first proposed an algorithm for the 
automation of perturbative calculation in lattice gauge theories \cite{lw1986}. 
Computer algebra tools for analytic calculations in LPT were 
developed in \cite{alles1993} and \cite{cap1995}, but it was not until 
twenty years after L\"uscher and Weisz's proposal that full automation started 
to receive greater attention within the lattice community. Hart \emph{et al.} 
extended the L\"uscher-Weisz algorithm to fermions in \cite{hart2005} and 
similar ideas were independently developed for the Schr\"odinger functional 
\cite{sf2005}. In the last three years two new ALPT routines have been 
developed, \pastor{} \cite{hesse2011} and \physycal{} \cite{lehner2012}, and 
these, in conjunction with \hipsrc, will be the focus of this section.

\subsection{\hipsrc} 

\hippy{} and \hpsrc{} are a suite of routines that automate
perturbative calculations for lattice QCD. \hippy{} is an implementation of the 
L\"uscher-Weisz algorithm in \texttt{Python}, extended to include the fermion 
sector. \hippy{} encodes the resulting Feynman rules as ``vertex files'', which 
are read into \hpsrc, a set of \texttt{FORTRAN} modules that evaluate the 
relevant Feynman diagrams numerically. 

This process is not fully automated, in the sense of Figure \ref{fig:auto}, in 
that the user must construct the desired Feynman diagrams by hand, but this 
procedure is conceptually identical to the standard application of the Feynman 
rules in continuum QCD. Modules within \hpsrc{} define functions that 
correspond to each vertex and propagator, including fermions, ghosts and gauge 
vertices up to five-gluon interactions. Feynman diagrams are constructed from 
these functions in the usual manner and then evaluated numerically via 
\texttt{VEGAS} \cite{lepage80} or exact mode summation (for finite volume 
lattices).

\paragraph{The L\"uscher-Weisz algorithm in \hipsrc} To illustrate in more 
detail how the process of deriving Feynman rules may be automated, I will 
briefly outline the L\"uscher-Weisz algorithm, as implemented in \hippy{}, and 
restrict the discussion to the gauge sector for brevity. The extension to 
fermions is fairly straightforward and documented in \cite{hart2005}.

At weak coupling we may represent the link 
variables as exponentials of the gauge fields. In general the gauge action 
is built from gauge invariant, \emph{i.e.}~closed, loops of link variables that 
can be represented in momentum space as
\begin{equation}
L(x,y;U) = \sum_r 
\frac{(ag)^r}{r!}\sum_{k_1,\mu_1,a_1}\ldots\sum_{k_r,\mu_r,a_r}\widetilde{A}_{
\mu_1}^{a_1}(k_1)\ldots\widetilde{A}_{\mu_r}^{a_r}(k_r) \times 
V_r(k_1,\mu_1,a_1;\ldots;k_r,\mu_r,a_r),
\end{equation}
where the $V_r$ are referred to as ``vertex functions''. These vertex 
functions can be further decomposed into a colour matrix, $C_r$, 
and a ``reduced vertex'', $Y_r$, that depends only on the momenta and positions 
of the links. This reduced vertex can be written as a product of exponentials 
\begin{equation}
V_r(k_i,\mu_i,a_i) = 
C_r(a_i)Y_r\left(k_i,\mu_i\right)=
C_r(a_i)\sum_{n=1}^{n_r}f_n\exp\left[\frac{i}{2}\left(k_1\cdot 
v_1^{(n)} + \ldots +
k_r\cdot v_r^{(n)}\right)\right],
\end{equation}
where $i \in \{1,\ldots, r\}$. Here the amplitudes associated with 
each term are denoted $f_n$ and the $v^{(n)}$ are the locations of each of the 
$r$ factors of the gauge potential. Therefore, for each combination of $r$ 
Lorentz vertices, there are $n_r$ terms, each with a corresponding amplitude 
$f_n$ and location $v_r^{(n)}$. Written in this form, the reduced vertices are 
encoded in ordered lists of ``entities'', $E = 
\left(\mu_1,\ldots,\mu_r;x,y;v_1,\ldots,v_r;f\right)$, which are stored in the 
vertex file generated by \hippy.

The Feynman rules, each written to a separate vertex file, are 
constructed by repeatedly applying the convolution formulae derived in 
\cite{morningstar1993}. So, for example, for an action of the form ${\cal 
O}_A{\cal O}_B$, the Feynman rule for, say, the three-gluon vertex is the 
convolution of the Feynman rules for three gluons chosen from operator ${\cal 
O}_A$ with no gluons from ${\cal O}_B$, plus the corresponding convolution for 
two gluons from ${\cal O}_A$ with one from ${\cal O}_B$ and so on.

The colour structure is completely encoded in the matrices $C_r$. For the 
\hipsrc{} package, this is handled in the \hpsrc{} routines
when the Feynman diagrams are constructed. Derivatives are obtained by 
algebraic manipulation of the reduced vertices, which ensures that no 
numerical derivatives are required and that the same vertex files can be used 
for both periodic and twisted boundary conditions. Within \hpsrc{}, analytic 
derivatives are implemented using the derived \texttt{taylor} 
type, defined as part of the \texttt{TaylUR} package for \texttt{FORTRAN}
\cite{gmvh2010}.

\subsection{Two recent calculations with \hipsrc}

A number of calculations have now been carried out using \hipsrc{} (see, 
for example, \cite{hart2004,dowdall2012,monahan2012}) and the code has been 
extensively tested. I would like to highlight two recent examples. For other 
applications, please see the review article \cite{monahan2012}.

\paragraph{Background field gauge in \hipsrc} The first of these developments
is the implementation of background field gauge in \hipsrc. Background field 
gauge (BFG) is a well established tool for gauge theories that has a number of 
advantages for calculations involving effective theories such as NRQCD.

Current lattice spacings are too coarse to directly incorporate $b$ quarks at 
their physical mass (whilst keeping lattice volumes large enough to avoid 
sizable finite volume effects). Traditionally this problem has been tackled by 
using an effective theory such as NRQCD or heavy quark effective theory (HQET) 
or with a relativistic heavy quark (RHQ) action that interpolates smoothly 
between relativistic and nonrelativistic limits. More recently, 
relativistic calculations that are extrapolated up to the physical 
$b$ mass have been developed \cite{hhisq2010}. Despite many advances, going to 
finer lattices is computationally very expensive and it is beneficial to retain 
an array of approaches to a problem to help identify and eliminate sources of 
systematic bias in lattice results. NRQCD and HQET remain an important 
ingredient in heavy quark physics on the lattice.

As part of the HPQCD Collaboration's programme of precision heavy quark 
physics, ALPT has been used to calculate the improvement coefficients for 
highly-improved NRQCD. The NRQCD action, which may be written as
\begin{equation}
S = \sum_{x,t} 
\Psi^\dagger(x,t)\Big[\Psi(x,t) - \left(1-\frac{a\delta 
H}{2}\right)\left(1-\frac{aH_0}{2n}\right)^nU_4^\dagger\left(1-\frac{aH_0}{2n}
\right)^n\left(1-\frac{a\delta H}{2}\right)\Psi(x,t-a)\Big],
\end{equation}
includes both the leading nonrelativistic kinetic energy term, $H_0 = 
\nabla^2/(2am_Q)$ and corrections:
\begin{align}
a \delta H = {} &
-c_1\frac{(\Delta^{(2)})^2}{8(am_Q)^3} +c_2\frac{ig}{8(am_Q)^2}
\big(\nabla\cdot \widetilde{E} - \widetilde{E}\cdot 
\nabla\big) -
c_3\frac{g}{8(am_Q)^2}\sigma\cdot\big(\widetilde{\nabla}\times 
\widetilde{E} - \widetilde{E}\times
\widetilde{\nabla} \big) \nonumber \\
{} & - c_4 \frac{g}{2am_Q}\sigma\cdot 
\widetilde{B} + c_5 
\frac{a^2\Delta^{(4)}}{24am_Q} - 
c_6\frac{a(\Delta^{(2)})^2}{16n(am_Q)^2} . 
\end{align}
The coefficients of these corrections are normalised to unity at 
tree-level, but radiative corrections alter their value. These 
radiative corrections can be determined at one-loop using ALPT and the first 
results, for the leading relativistic and discretisation corrections ($c_1$, 
$c_5$ and $c_6$), were presented in \cite{dowdall2012}.

More recently, the one-loop radiative corrections to the chromo-magnetic, 
$c_4$, and Darwin, $c_2$, terms were calculated using BFG \cite{hammant2011}. 
Using BFG greatly simplifies the matching calculation by
\begin{itemize}
\setlength{\itemsep}{-3pt}
\item ensuring that only 
gauge-covariant $D>4$ operators appear in the NRQCD effective action;
\item rendering the relevant one-particle irreducible vertex function 
ultraviolet finite; and
\item constraining renormalisation parameters via an extra background field 
Ward identity.
\end{itemize}
The first of these guarantees that we need only include gauge-invariant 
operators in the action and the second allows us to match lattice NRQCD 
directly to QCD. The third simplifies the calculation by reducing the number of 
renormalisation parameters that must be calculated.

The effectiveness of this improvement programme is demonstrated by the 
improvement in the lattice NRQCD result for the bottomonium hyperfine 
splitting, $(M_\Upsilon(1S) - M_{\eta_b}(1S))$, when radiative corrections are 
included in the NRQCD action \cite{dowdall2013}. An earlier determination by 
the HPQCD Collaboration obtained 56(2) 
MeV for the hyperfine splitting, in disagreement with the world experimental 
average of 62.2(3.2) MeV \cite{pdg2013}. Incorporating radiative corrections 
(amongst other improvements), HPQCD Collaboration's new result is 62.8(6.7) MeV, 
now in good agreement with experimental data.

\paragraph{The $b$ quark mass from ALPT and weak-coupling computations} The 
HPQCD Collaboration's new determination of the $b$ quark mass from lattice 
NRQCD incorporated a mixed approach to determine perturbative results to 
next-to-leading order (\emph{i.e.}~two-loops).

Quark masses are conventionally expressed in the $\overline{MS}$ scheme, and 
for the $b$ quark, typically evaluated at a scale equal to the $b$ quark 
mass itself, $\overline{m}_b(\overline{m}_b)$. The $\overline{MS}$ scheme is
inherently perturbative and the $\overline{MS}$ mass must be related to 
nonperturbative quantities (determined on the lattice) via a two stage matching 
procedure with the pole mass as an intermediate step \cite{lee2013}. For NRQCD, 
one method is to use the heavy quark energy shift, $E_0$,
\begin{equation}
\overline{m}_b(\mu) = \frac{1}{2}Z_M^{\,-1}(\mu)\left[
M_\Upsilon^{\,\mathrm{expt}} - 
a^{-1}\left(aE_\Upsilon^{\,\mathrm{sim}} - 
2a E_0\right)\right].
\end{equation}
Here $Z_M$ is the renormalisation parameter relating the pole-mass to the
$\overline{MS}$ mass, known to three-loops \cite{mvr2000}, 
$M_\Upsilon^{\,\mathrm{expt}}$ is the experimental $\Upsilon$ meson mass 
(corrected for missing electromagnetic effects) and 
$E_\Upsilon^{\,\mathrm{sim}}$ is the ground state energy of the $\Upsilon$ 
meson determined from lattice NRQCD. The quantity 
$\left(aE_\Upsilon^{\,\mathrm{sim}} - 
2a E_0\right)$ represents the ``binding'' energy of the heavy quarks.

The HPQCD Collaboration's earlier result from NRQCD, 
$\overline{m}_b(\overline{m}_b) = 4.4(3)$ GeV, was dominated by the two-loop 
perturbative uncertainty in the energy shift, $E_0$ \cite{gray2005}. By 
including not only the two-loop contribution to the energy shift, but also the 
three-loop quenched contribution as well, the error was reduced to the one 
percent level: $\overline{m}_b(\overline{m}_b) = 4.166(43)$ GeV \cite{lee2013}.

In this calculation, the two-loop quenched contributions to the 
energy shift, which include all 27 gauge and ghost field diagrams and 
propagator insertions, were determined from weak coupling computations of 
lattice NRQCD in the quenched approximation. Results were obtained for a range 
of bare coupling values and the perturbative coefficients extracted from a 
simultaneous fit to a polynomial in the strong coupling constant and the inverse 
lattice size. The remaining fermionic diagrams were calculated using ALPT. 

This approach combined the advantage of the (relative) computational speed of 
quenched lattive calculations with the (relative) simplicity of calculating 
only four diagrams with ALPT. The decrease in overall uncertainty in the final 
result for the $b$ 
quark mass is largely due to the inclusion of the two-loop (and 
quenched three-loop) contributions to the heavy quark energy shift. Such 
improvement demonstrates the effectiveness of the mixed approach to 
higher-order perturbation theory and underscores the importance of ALPT for 
precision lattice calculations.

\subsection{\pastor} 
The ALPT package \pastor{} is a suite of routines based on the 
L\"uscher-Weisz algorithm and tailored to the 
Schr\"odinger functional (SF) scheme. Developed by Dirk Hesse, the details of 
the implementation are discussed in \cite{hesse2012} and initial 
calculations outlined in \cite{hesse2011}.

\begin{figure}
\begin{center}
\frame{\includegraphics[width=0.7\textwidth,keepaspectratio=true]{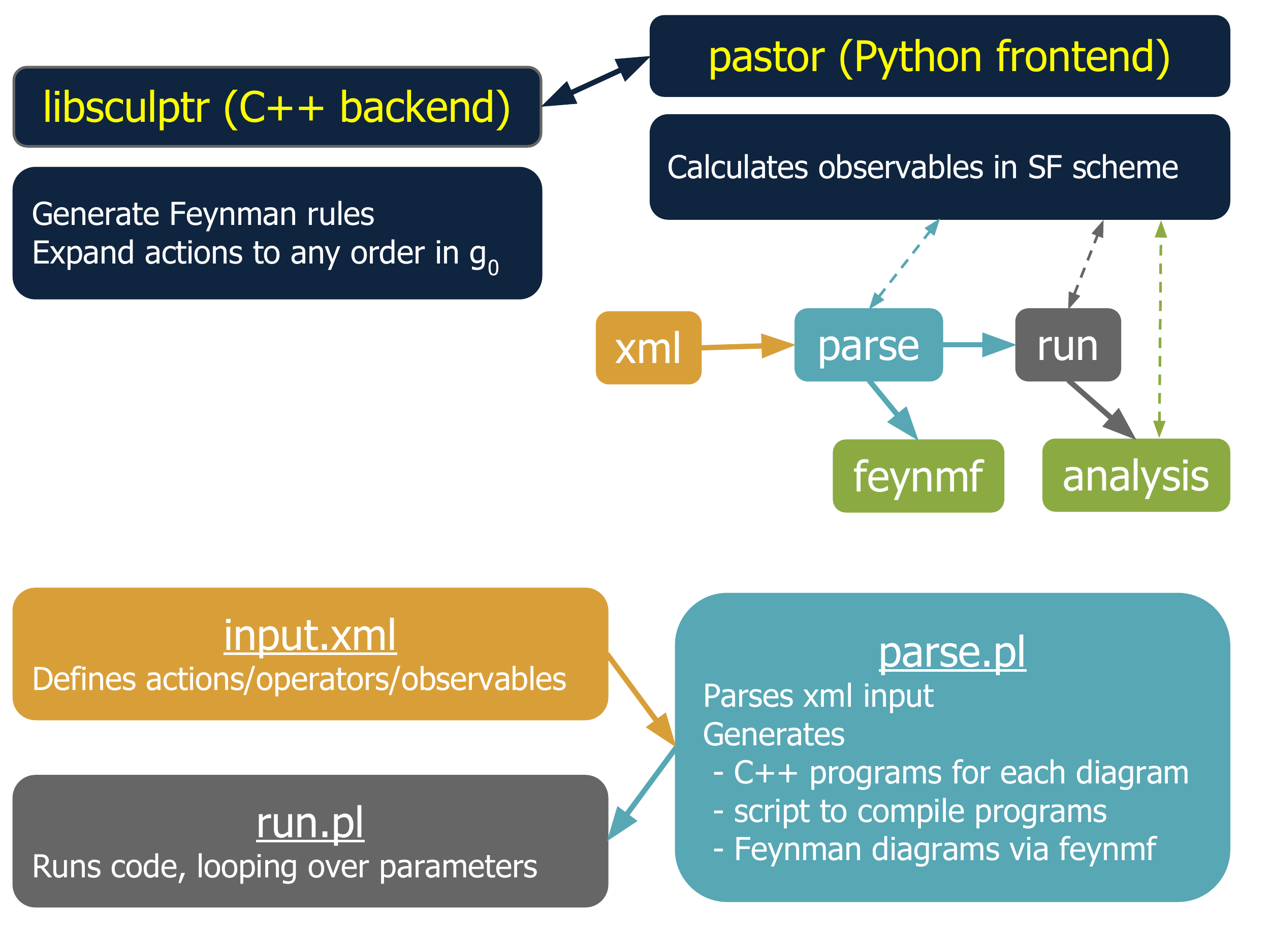}
}
\end{center}
\caption{Schematic representation of the structure of \pastor. See accompanying 
text for details.\label{fig:pastor}}
\end{figure}

I illustrate the structure of \pastor{} in Figure 
\ref{fig:pastor}. Vertices are generated by \texttt{libsculptr}, the 
\texttt{C++} backend of \pastor. Feynman diagrams are generated and evaluated by 
a set of \texttt{Python} routines: \texttt{parse.py}, \texttt{run.py}, and 
\texttt{analysis.py}. The user specifies all inputs (action, parameters, 
observable \emph{etc.}) in an \texttt{xml} file, which is parsed by 
\texttt{parse.py} that, in turn, generates efficient \texttt{C++} code to 
evaluate the relevant diagrams. This \texttt{C++} code is called by 
\texttt{run.py}{} and written in a format that is easy to parse with the 
analysis script \texttt{analysis.py}. Feynman diagrams can be automatically 
created using the \texttt{feynmf} package in \LaTeX.

\subsection{Matching HQET and QCD with \pastor}

Lattice HQET is an approach for heavy quarks on the lattice that is
suitable for heavy-light mesons (for an overview, see, for example, 
\cite{sommer2010}) and is currently being used by, among others, the ALPHA 
Collaboration. One of the principle differences from NRQCD is that higher order 
terms are 
included as operator insertions, rather than directly in the action. This 
ensures that HQET is renormalisable. The HQET Lagrangian at ${\cal 
O}(1/m_Q)$ may be written as
\begin{equation}
{\cal L}_{\text{HQET}} = {\cal L}_{\text{stat}} - 
\left(\omega_{\text{kin}}{\cal L}_{\text{kin}} + \omega_{\text{spin}}{\cal 
L}_{\text{spin}}\right) + {\cal O}(1/m_Q^2), 
\end{equation}
where ${\cal L}_{\text{stat}}$ is the leading order static 
(\emph{i.e.}~infinite mass limit) Lagrangian and ${\cal L}_{\text{kin}}$ 
and ${\cal L}_{\text{spin}}$ are the relativistic and chromo-magnetic 
corrections respectively. Correlation functions of operators are given by 
\begin{equation}
\langle O \rangle_{\text{HQET}} = \langle 
O\rangle_{\text{stat}}+\omega_{\text{kin}}\sum_x \langle O{\cal 
L}_{\text{kin}}(x)\rangle_{\text{stat}}+\omega_{\text{spin}}\sum_x \langle 
O{\cal 
L}_{\text{spin}}(x)\rangle_{\text{stat}}.
\end{equation}

Lattice HQET must be matched to continuum 
QCD to extract physically meaningful results and this matching must be 
nonperturbative to avoid renormalon ambiguities. Lattice HQET 
observables are matched to continuum QCD via
\begin{equation}
\phi_i^{\text{QCD}}(L,z,a=0) = \phi_i^{\text{HQET}}(L,\;z\;,a; \omega(z,a))  = 
\phi_i^{\text{stat}}(L,a) + \phi_{ij}^{(1/z)}(L,a)\omega_j(z,a).
\end{equation}
Here $L$ is the lattice size, $a$ the lattice spacing, and $z$ is the 
dimensionless mass $z=\overline{m}(L)L$.

Achieving the one percent precision 
necessary for precision $B$ physics
requires matching at ${\cal O}(1/m_Q)$. At this order, there are 
nineteen parameters that must be fixed. Computationally, this is a very 
expensive task and to guide the process ALPT was used to investigate the 
$ {\cal O}(1/m_Q^2)$-dependence of different operators and test the quality of 
observables for matching \cite{hesse2011,korcyl2013}.

\begin{figure}
\centering
\begin{minipage}{.5\textwidth}
\centering
\includegraphics[width=\textwidth,keepaspectratio=true]{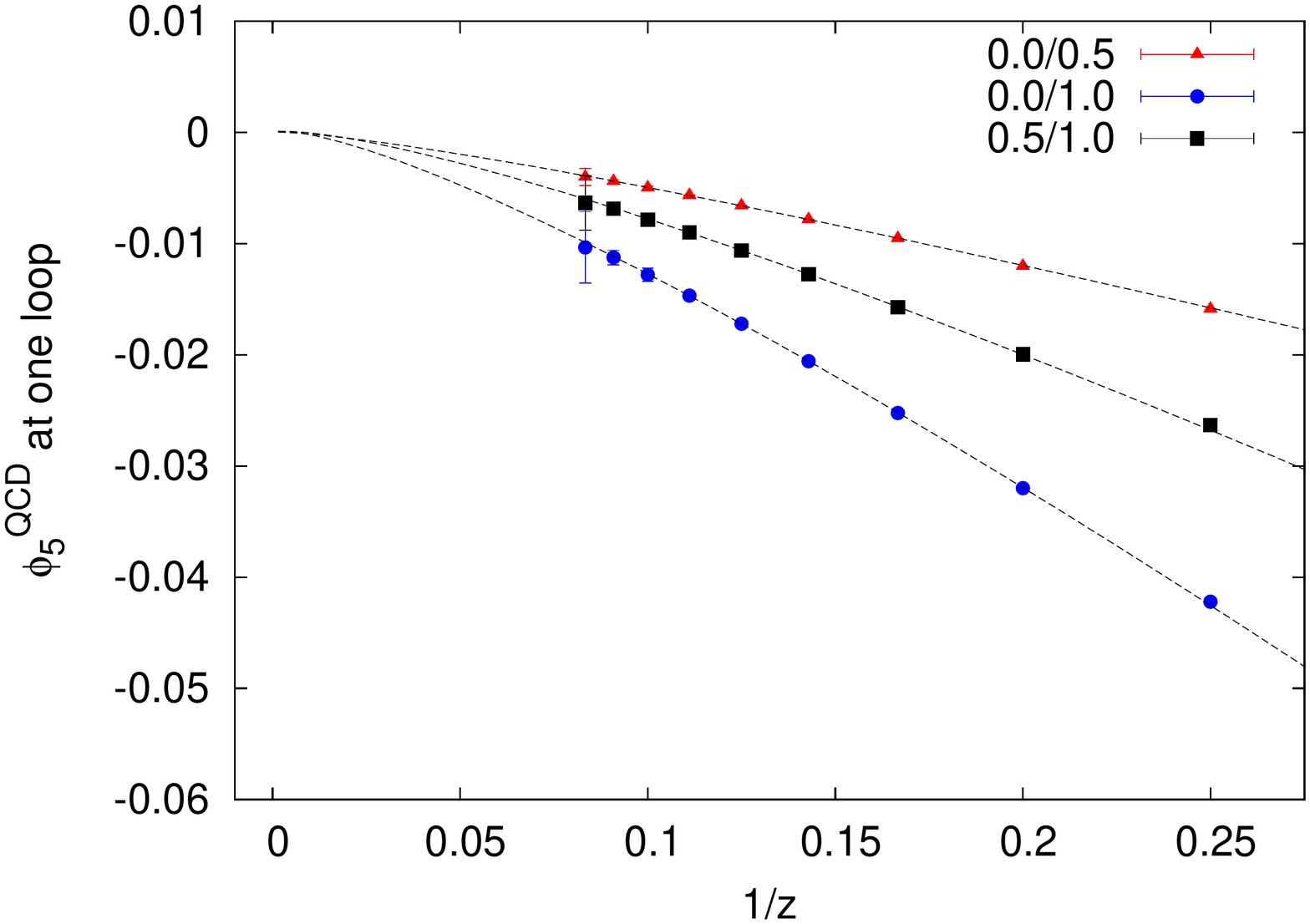}
\end{minipage}%
\begin{minipage}{.5\textwidth}
\centering
\includegraphics[width=\textwidth,keepaspectratio=true]{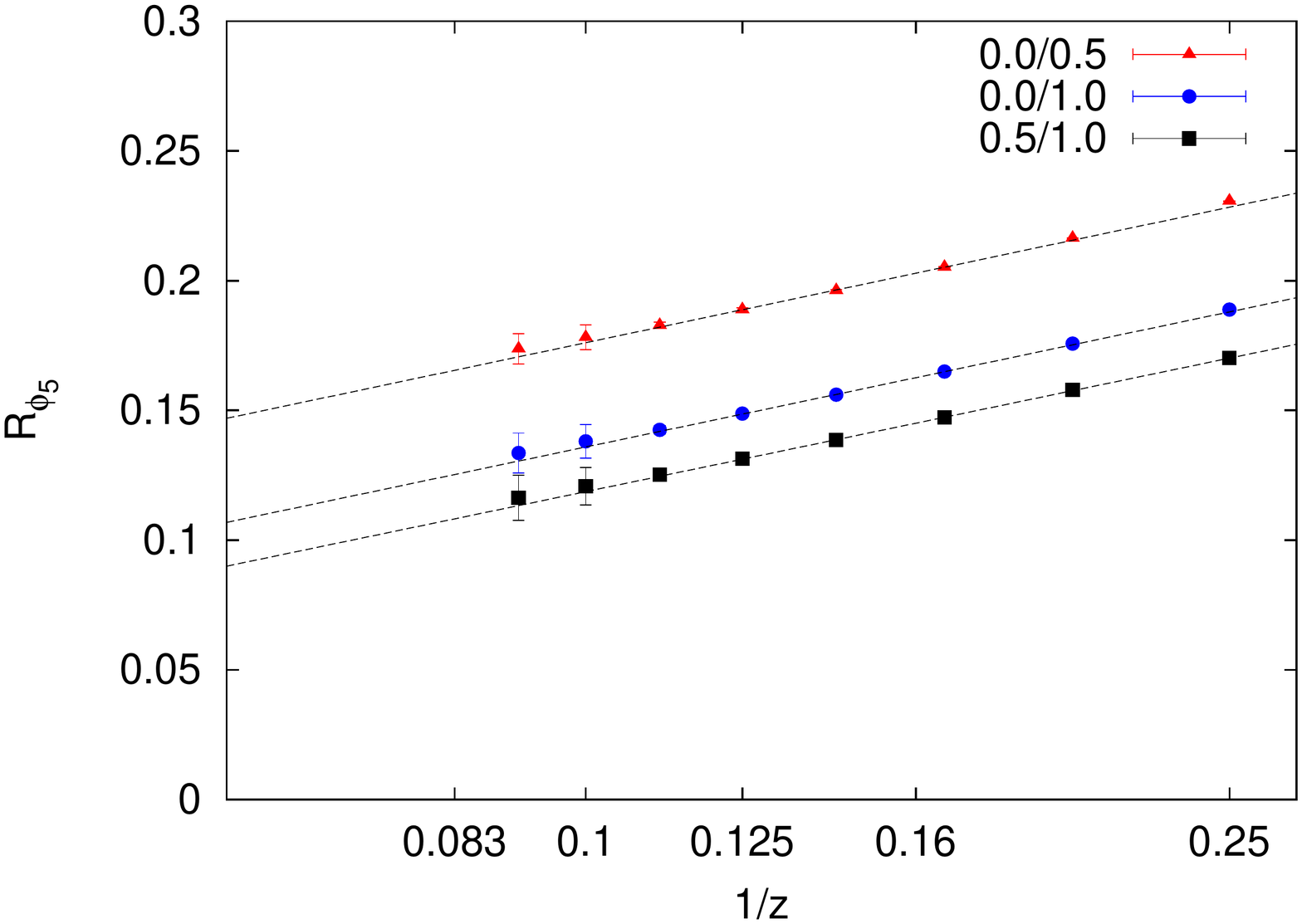}
\end{minipage}
\caption{Mass dependence of the HQET observable $\phi_5$. For more details see 
the accommpanying text and Ref.~\cite{korcyl2013}.\label{fig:hqet}}
\end{figure}

The mass dependence can be studied by examining the ratio
\begin{equation*}
R_{\phi_i} = \frac{\phi^{\text{QCD},(1)}_i(z) - 
\phi^{\text{stat},(1)}_i}{\phi^{\text{QCD},(0)}_i(z) - 
\phi^{\text{stat},(0)}_i}
\end{equation*}
and then plotting the result on a linear-log plot. Deviations from linear 
behaviour indicate $1/z^2$ contributions and the slope gives the coefficient of 
subleading logarithmic behaviour. An example of such a plot is given in Figure 
\ref{fig:hqet}, which shows the mass dependence of the observable $\phi_5$ 
(for details and definitions, see \cite{korcyl2013}). The righthand plot 
demonstrates that there is little contamination by $1/z^2$ contributions in 
$\phi_5$, which indicates that it likely to be a useful observable in 
nonperturbative matching calculations for lattice HQET.

Matching conditions for all nineteen parameters have now been investigated 
\cite{korcyl2013}, a task that has provided considerable insight into the best 
choice of operators for the final matching procedure. Without ALPT, this simply 
would not have been feasible in a reasonable timescale.

\subsection{\physycal}

The final package that I would like to discuss is \physycal{}, which 
takes a completely different approach to automating LPT. 
\physycal{} is a computer algebra system (CAS), \emph{i.e.}~a 
software package for symbolic manipulation of mathematical 
expressions, developed by Christoph Lehner \cite{lehner2012}.

\begin{figure}
\begin{center}
\frame{\includegraphics[width=0.7\textwidth,keepaspectratio=true]{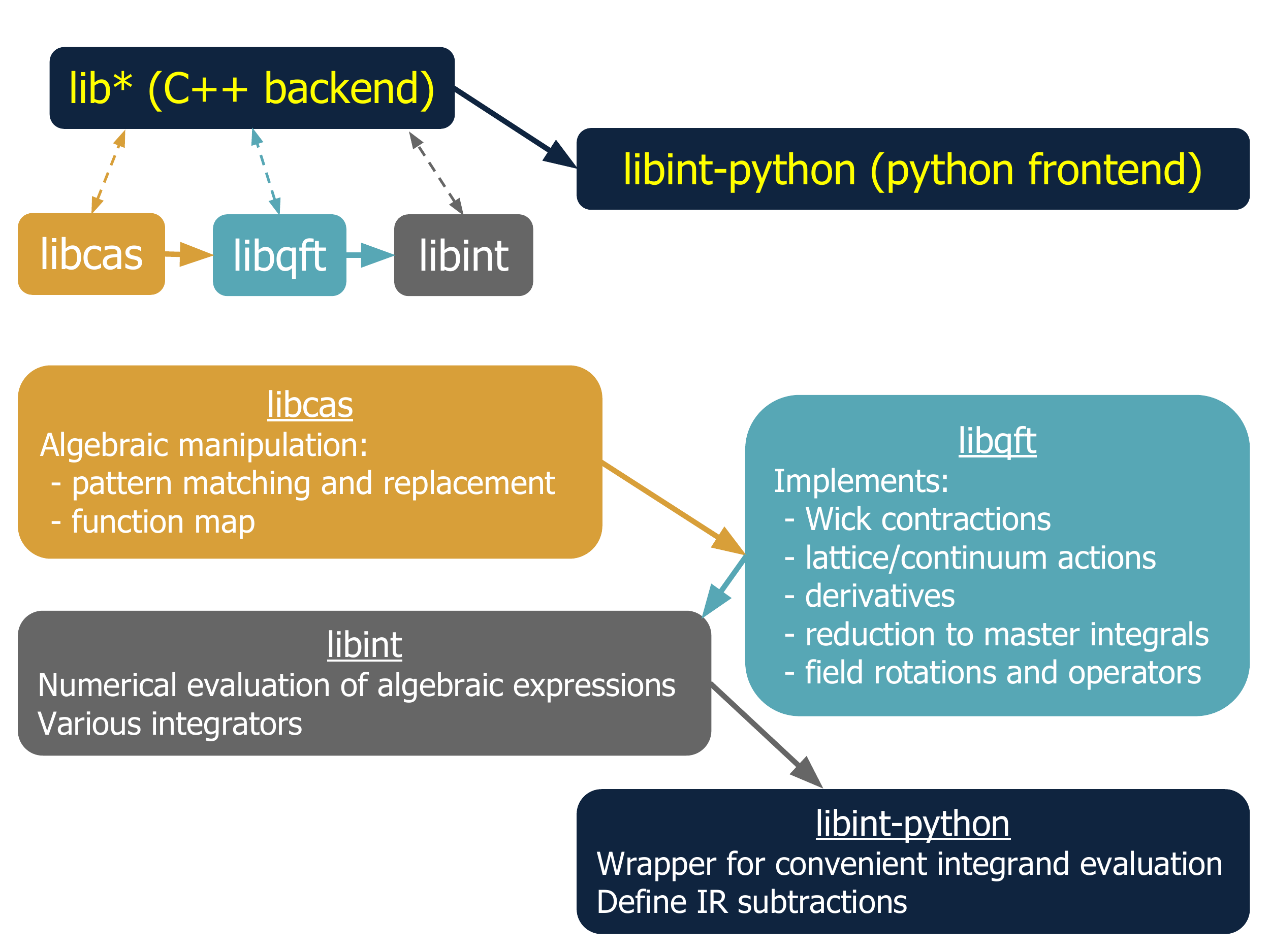}}
\end{center}
\caption{Schematic representation of the structure of \physycal. See 
accompanying text for details.\label{fig:cas}}
\end{figure}

The structure of \physycal{} is outlined in Figure \ref{fig:cas}. The heavy 
lifting is done by three \texttt{C++} libraries: \texttt{libcas}, the CAS; 
\texttt{libqft}, a quantum field theory library; and \texttt{libint}, a library 
for numerical evaluation of algebraic expressions.

The CAS library \texttt{libcas} parses algebraic expressions from text and 
\texttt{xml} input files and simplifies these expressions. \texttt{libqft} 
performs all manipulations associated with typical perturbative 
calculations: deriving vertices from actions and operators; carrying out Wick 
contractions; and even computing loop integrals in dimensional regularisation 
via the Passarino-Veltmann reduction. Finally, the loop integrals are evaluated 
numerically, using a choice of integrators, with \texttt{libint}.

\subsection{Matching relativistic heavy quarks with \physycal}

In the Columbia formulation, used by the RBC/UKQCD Collaboration for its $B$ 
physics programme, the RHQ action may be written as
\begin{equation}
{\cal S} = \sum_x 
\overline{Q}(x)\bigg[\left(\gamma_0D_0 - 
\frac{D_0^2}{2}\right) + \zeta \sum_{i=1}^3\left(\gamma_iD_i - 
\frac{D_i^2}{2}\right) \nonumber + m_0 
+\frac{ic_P}{4}\sum_{\mu,\nu=0}^3\sigma_{\mu\nu}F_{\mu\nu}\bigg]Q(x),
\end{equation}
where the parameters $\zeta$, $m_0$, and $c_P$ can be tuned to remove 
discretisation effects \cite{lehner2012}. Applying a field rotation to the 
heavy quark fields, $Q(x)\rightarrow Q^\prime(x) = Q(x) + 
d_1\sum_{i=1}^3\gamma_iD_iQ(x)$,
the rotated fields may be matched to continuum fields. Since this rotation 
leaves the mass spectrum unchanged, the three parameters $\zeta$, $m_0$, and 
$c_P$ can be tuned without knowledge of $d_1$.

Matching the RHQ quark two-point function to continuum QCD 
allows one to tune $m_0$ via the pole mass and $\zeta$ using the dispersion 
relation. The parameter $c_P$ is extracted by matching the quark-quark-gluon 
three-point function in the on-shell limit. There are 37 Feynman diagrams 
required to match $c_P$ to one-loop, 
which is clear evidence that automating the procedure dramatically simplifies 
the calculation.

\begin{figure}
\centering
\hspace*{-3pt}\begin{minipage}{.5\textwidth}
\centering
\includegraphics[width=\textwidth,keepaspectratio=true]{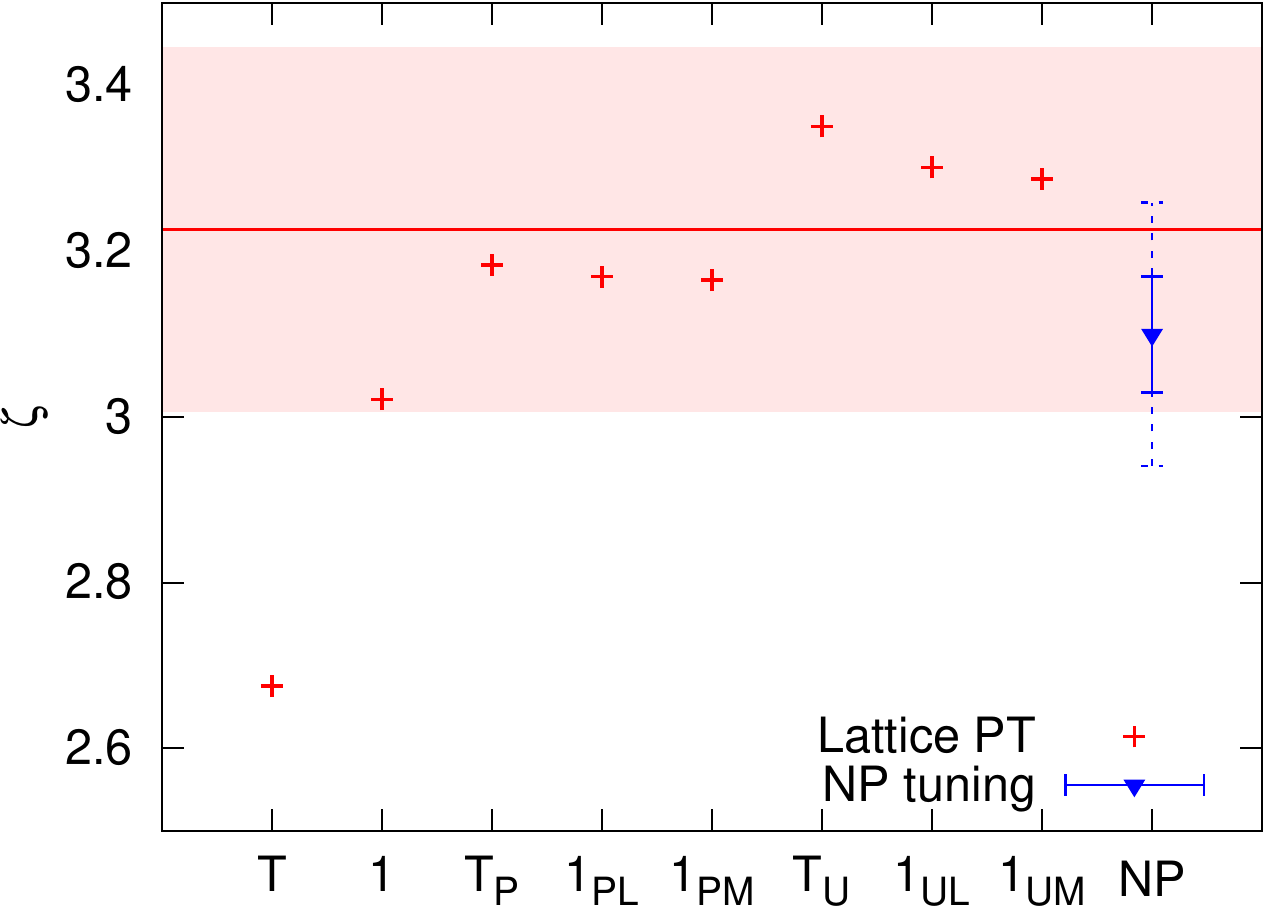}
\end{minipage}%
\hspace*{3pt}\begin{minipage}{.5\textwidth}
\centering
\includegraphics[width=\textwidth,keepaspectratio=true]{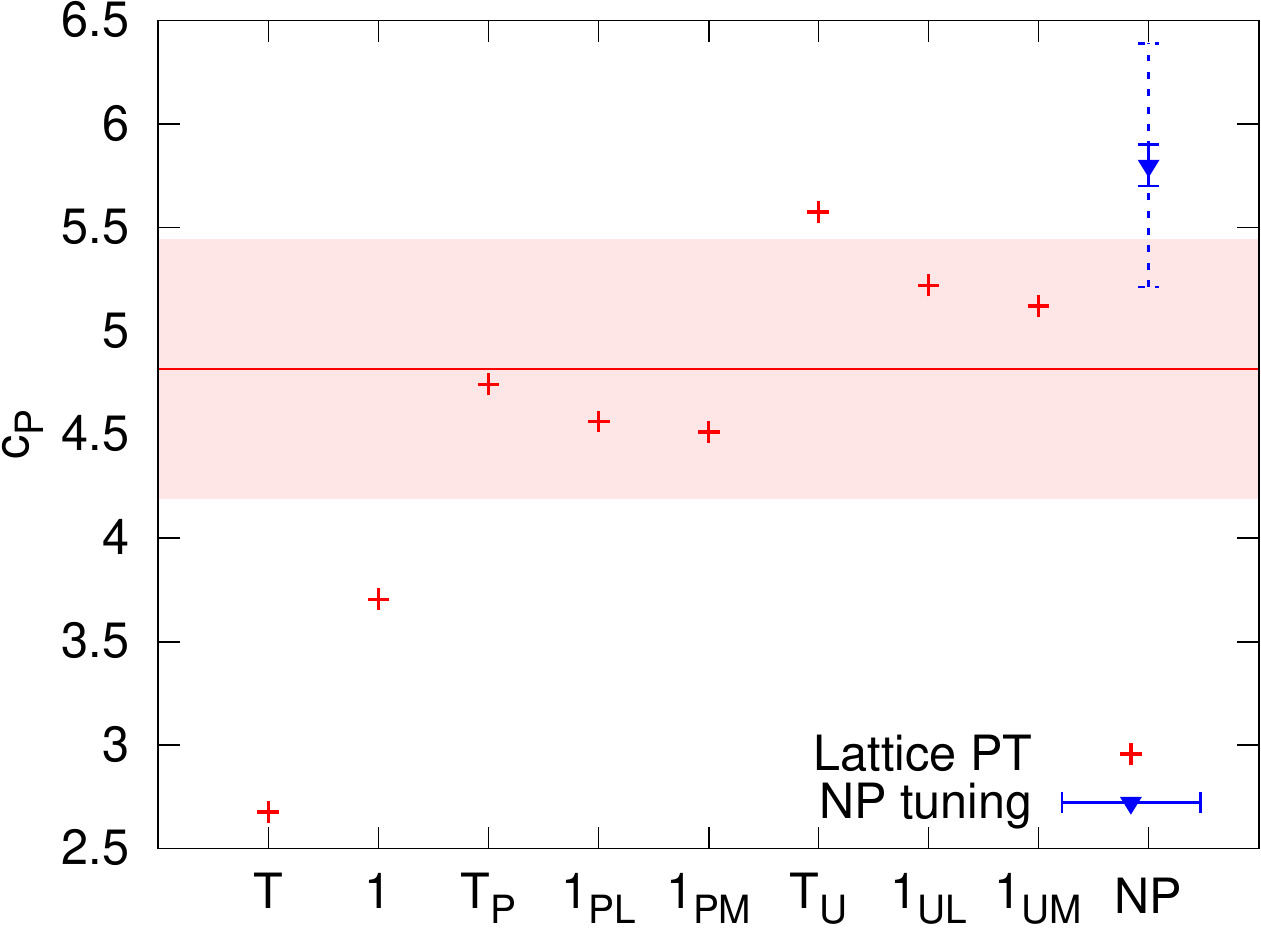}
\end{minipage}
\caption{Perturbative and nonperturbative results for the RHQ parameters 
$\zeta$ (on the left) and $c_P$ (on the right). The different points 
correspond to different choices of perturbative coefficient, improvement 
scheme and coupling constant. For full details see \cite{lehner2012}. The 
nonperturbative result is plotted in blue for comparison.\label{fig:rhq}}
\end{figure}

\physycal was used to test the matching framework for the RHQ quarks, by 
calculating the one-loop coefficients of $\zeta$, $m_0$, and 
$c_P$. Sample data for both $\zeta$ and $c_P$ are presented in Figure 
\ref{fig:rhq}.
These perturbative results were compared 
to the nonperturbatively tuned values, serving as a cross-check of both 
methods. It is clear from Figure \ref{fig:rhq} that perturbative and 
nonperturbative results agree within errors. Nonperturbative tuning is 
computationally expensive and currently not available for all operators relevant 
to the RBC/UKQCD $B$ physics programme. ALPT, implemented using the \physycal{} 
package, offers an alternative method for determining renormalisation parameters 
and tuning the RHQ action that is likely to prove highly useful in the near 
future. 

\section{Numerical Stochastic Perturbation Theory}
\label{sec:nspt}

Numerical stochastic perturbation theory (NSPT) lies on a rather different 
branch of the evolutionary tree of approaches to LPT than the automated 
routines just discussed. Proposed in 
\cite{direnzo1994}, NSPT is a numerical application of stochastic quantisation, 
particularly suited to high-order calculations (for a good review see, for 
example \cite{direnzo2004}).

Stochastic quantisation proceeds by introducing an extra degree of freedom, the 
stochastic time $\tau$, in which the fields evolve according to a Langevin 
equation with Gaussian noise.

The central claim of stochastic quantisation is that stochastic 
averages, taken over an ensemble of statistical noise sources $\eta$, converge 
to the original functional integral expectation values of the theory in the 
infinite stochastic time limit. That is,
\begin{equation}
\left\langle O[\phi(x,\tau)] \right\rangle_\eta 
\stackrel{\tau\rightarrow \infty}{\longrightarrow} \left \langle O[\phi(x)] 
\right\rangle,
\end{equation}
where the stochastic average on the left is over the Gaussian noise and the 
right hand side is the usual field theoretic expectation value.

Stochastic perturbation theory is obtained by decomposing 
the action into a free field component and an interacting contribution that is 
a function of the bare coupling constant. One then iteratively solves the 
Langevin equation for the full theory as a series in the coupling 
constant. In the infinite stochastic time limit, the 
original perturbative expansion is recovered. NSPT is the numerical 
implementation of this procedure.

In the interests of brevity I have skipped over many interesting subtleties, 
such as incorporating fermions, stochastic gauge fixing and navigating 
numerical instabilities, that are beyond the scope of these proceedings and 
refer the reader to the literature (see, for example, \cite{lptorg}).

\subsection{Recent calculations with NSPT}

There have been a wide range of NSPT calculations carried out since its 
inception and I discuss here only the most recent results.

\begin{figure}
\begin{center}
\includegraphics[width=0.55\textwidth,keepaspectratio=true]{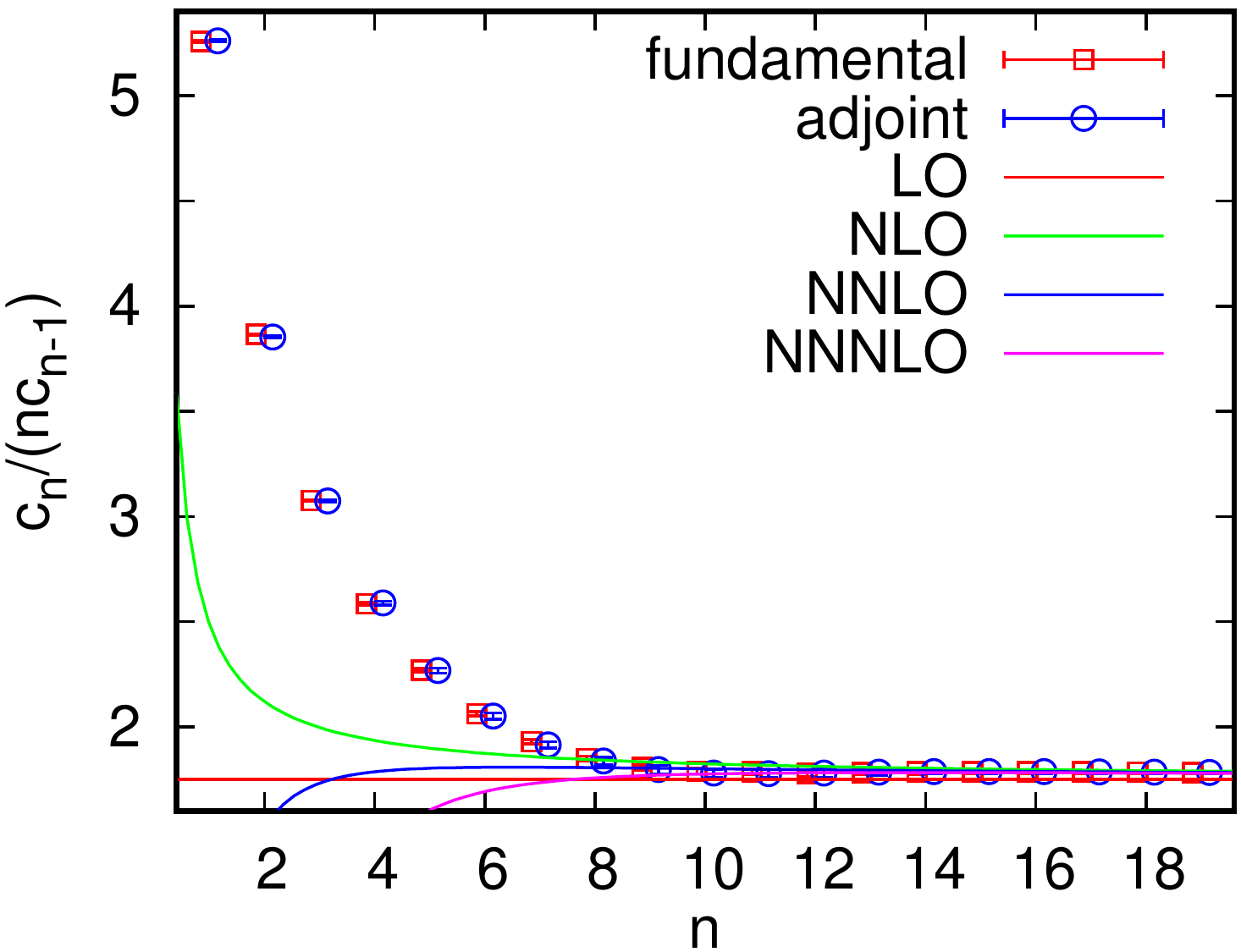}
\end{center}
\caption{Plot showing the $n!$ growth of coefficients in the self-energy of a 
static field source in $SU(3)$ gauge theory. Results for the fundamental 
representation are shown in red and for the adjoint representation in blue. 
Plotted lines illustrate the expected behaviour from perturbation theory 
at leading order (LO, in red) 
up to next-to-next-to-leading order (NNLO, in blue) and an estimate of the 
next-to-next-to-next-to-leading order (NNNLO, in pink). Plot simplified from 
\cite{bali2013}.\label{fig:pline}}
\end{figure}

\paragraph{Evidence for renormalons} In \cite{bali2013} strong evidence for the 
existence of renormalons was obtained by studing the Polyakov loop
at very high orders, up to ${\cal O}(\alpha^{20})$. The main results are 
summarised in Figure \ref{fig:pline}, where the $n!$ growth of coefficients 
characteristic of renormalons is evident. In fact, the normalisation of the 
leading renormalon of the heavy quark
pole mass, which is related to the Polyakov line, was found to differ
from zero by ten standard deviations, providing compelling numerical evidence 
for the existence of the leading infrared renormalon.

\begin{figure}
\begin{center}
\includegraphics[width=0.9\textwidth,keepaspectratio=true]{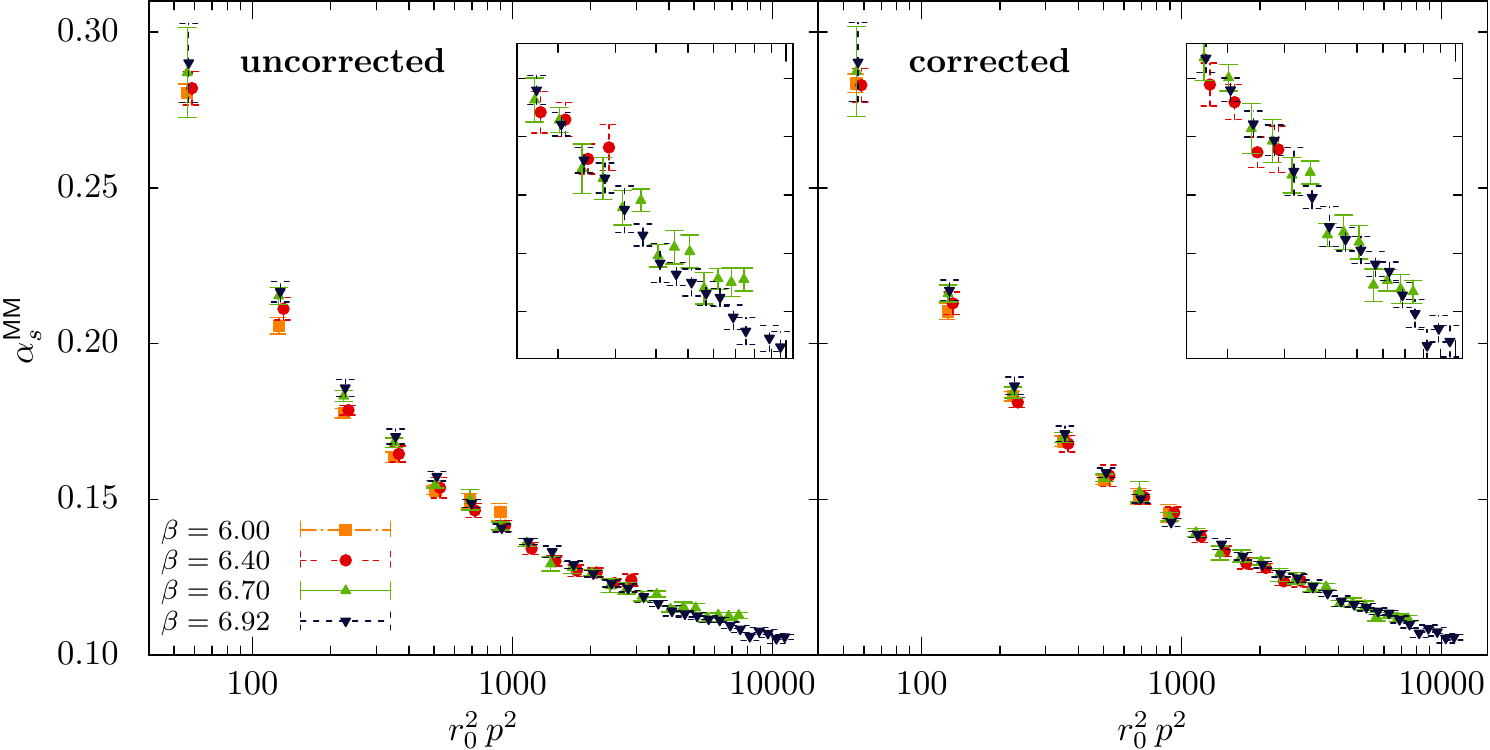}
\end{center}
\caption{Coupling constant in the ``Minimal MOM'' 
scheme \cite{sms2009} for quenched QCD at fixed physical volume, before (left) 
and after (right) correcting for discretisation effects \cite{simeth2013}. The 
corrections were calculated using NSPT at three loops and results are plotted 
at four different values of $\beta$.\label{fig:alphas}}
\end{figure}

\paragraph{Discretisation errors in the MiniMom scheme} NSPT has recently been 
used to investigate and correct discretisation errors in the strong coupling 
constant, defined in the ``Minimal MOM'' scheme \cite{sms2009}. The Minimal MOM 
coupling is renormalisation-scale invariant, but on the lattice this invariance 
is slightly broken by lattice artefacts at ${\cal O}(a^2)$. These effects are 
seen in the plot on the lefthand side of Figure \ref{fig:alphas}, where the 
points do not lie perfectly on top of each other. These artefacts are mainly 
due to loop contributions to the ghost and gluon propagators used to extract 
the coupling. Such discretisation effects become more important for unquenched 
lattice calculations, where very fine lattices are 
prohibitively expensive.

Previous attempts to calculate the discretisation corrections at one loop in 
standard LPT proved insufficient to remove the lattice artefacts completely
\cite{sternbeck2012} and higher orders were required. The righthand side of 
Figure \ref{fig:alphas} shows the effect of the three-loop 
corrections calculated with NSPT: the points at different spatial momenta now 
lie almost on top of each other. Although only a preliminary study in the 
quenched approximation, these results demonstrate the effectiveness of NSPT as 
a tool for correcting lattice artefacts.

\paragraph{NSPT and the gradient flow} The gradient flow is a technique similar 
in spirit to stochastic quantisation, but without the inclusion of a stochastic 
noise term in the evolution equation. The evolution in the flow time, $\tau$, 
is governed by a deterministic partial differential equation that drives the 
fields to a stationary point of the action, a process that corresponds to 
repeatedly smearing the fields. For a review of new developments involving the 
gradient flow, see \cite{luescher2013}.

\begin{figure}

\begin{center}
\includegraphics[width=0.8\textwidth,keepaspectratio=true]{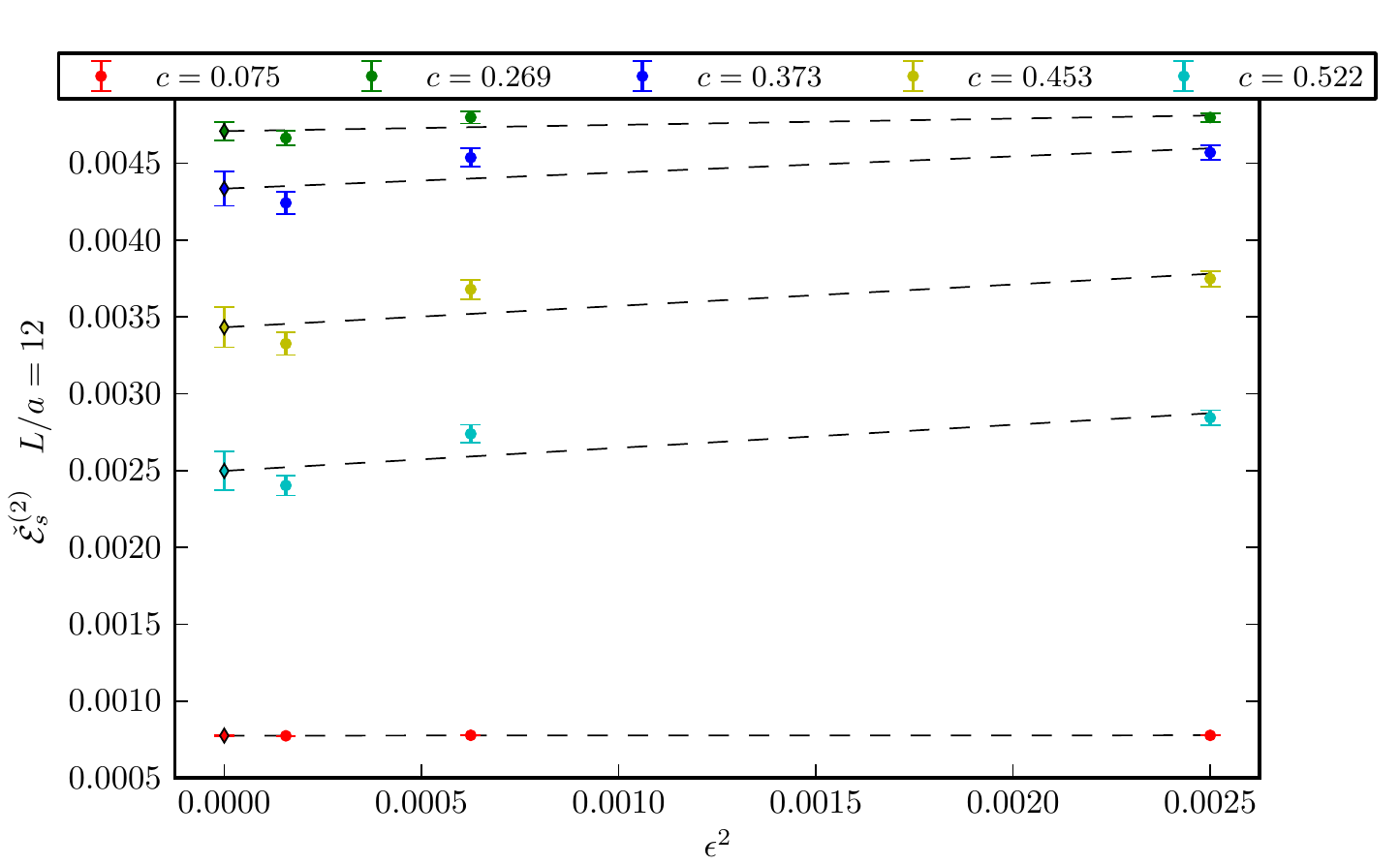}
\end{center}
\caption{A plot of the two-loop field strenth, ${\cal E}^{(2)}$, against 
stochastic time step size. Each choice of $c$, where $c^2 = 8\tau/L^2$ (see 
accompanying text), represents a different renormalisation 
scheme.\label{fig:nsptflow}}

\end{figure}

NSPT has recently been applied to the gradient flow method in 
the Schr\"odinger functional scheme \cite{mdb2013}. In particular,  
the gauge field strength, ${\cal E}$, was studied and the results of 
\cite{fr2013} for the normalisation of ${\cal E}$ were reproduced. The use of 
NSPT allowed this result to be extended to the next two orders. Figure 
\ref{fig:nsptflow} shows the field strength at two-loops, 
${\cal E}^{(2)}$, as function of the stochastic step size, $\epsilon$, and for 
five choices of flow time constant $c$. This constant relates the flow time, 
$\tau$, to the box size, $L$, via $c^2 = 8\tau/L^2$ and different choices 
represent different renormalisation schemes for the finite volume coupling 
constant defined through ${\cal E}$ \cite{fr2013}. The final errors in the 
two-loop field strength are only 3\%, suggesting that NSPT will be a promising 
tool for precision calculations in the Schr\"odinger functional scheme in the 
future (see also \cite{bhdr2013}).

\section{Summary}
\label{sec:conc}

As lattice calculations become more precise, the need to reduce all 
sources of systematic uncertainty grows more pressing. LPT plays an 
important role in reducing systematic uncertainties, from improving actions to 
calculating renormalisation parameters and mixing coefficients. As lattice 
actions become more sophisticated, automation becomes a necessity and the 
effectiveness of ALPT has been demonstrated in 
a wide variety of calculations over the last ten years. Even in theories 
requiring nonperturbative renormalisation, ALPT has a role to play as a guide to 
choosing suitable renormalisation conditions and as a test of nonperturbative 
results. 

In many 
calculations perturbative uncertainties dominate the error budgets and improved 
and higher-order calculations are required. ALPT, perhaps combined with weak 
coupling calculations, and NSPT will certainly be the only realistic methods 
for carrying out higher-order calculations in the future. With the 
advent of new automation routines, the use of ALPT is 
only set to grow.

\acknowledgments

I would like to thank Christoph Lehner, Dirk Hesse and Piotr Korcyl for 
many informative conversations about ALPT and all the authors who 
contributed their time and their plots for this plenary. Particular thanks to 
Chris Bouchard and Kelcie Ralph for reading a draft of the manuscript.

\end{document}